\documentclass[reprint,aps,prb,twocolumn,floats,showkeys,footinbib,superscriptaddress]{revtex4-1}
\usepackage[T1]{fontenc}
\usepackage{bm}
\usepackage{graphicx}
\usepackage{amssymb}
\usepackage{leftidx}
\usepackage{upgreek}
\usepackage{siunitx}
\usepackage{amsfonts}
\usepackage{amsmath} 
\usepackage[outdir=./]{epstopdf}
\usepackage{color}
\usepackage{float}
\usepackage{verbatim}
\usepackage{hyperref}
\usepackage{amssymb}
\usepackage{wasysym}

\begin{document}
\title{\large \textbf{Magnetic ion relaxation time distribution within a quantum well}} 

\def \FUW{Institute of Experimental Physics, Faculty of Physics, University
of Warsaw, ul. Pasteura 5, 02-093 Warsaw, Poland}

\author{A. \surname{{\L}opion}}\email[Corresponding author:]{Aleksandra.Lopion@fuw.edu.pl}\affiliation{\FUW}
\author{A. \surname{Bogucki}}\email[Corresponding author:]{Aleksander.Bogucki@fuw.edu.pl}\affiliation{\FUW}
\author{W. \surname{Kra\'{s}nicki}}\affiliation{\FUW}
\author{K.\,E. \surname{Po{\l}czy\'{n}ska}}\affiliation{\FUW}
\author{W. \surname{Pacuski}}\affiliation{\FUW}
\author{T. \surname{Kazimierczuk}}\affiliation{\FUW}
\author{A. \surname{Golnik}}\affiliation{\FUW}
\author{P. \surname{Kossacki}}\affiliation{\FUW}

\begin{abstract}
Time-resolved optically detected magnetic resonance (ODMR) is a valuable technique to study the local deformation of the crystal lattice around magnetic ion as well as the ion spin relaxation time. Here we utilize selective Mn-doping to additionally enhance the inherent locality of the ODMR technique. We present the time-resolved ODMR studies of single {(Cd,Mg)Te/(Cd,Mn)Te} quantum wells (QWs) with manganese ions located at different positions along the growth axis -- in the center or on the sides of the quantum well. We observe that spin-lattice relaxation of Mn$^{2+}$ significantly depends on the ion-carrier wavefunction overlap at low-magnetic fields. Interestingly, the effect is clearly observed in spite of very low carrier density, which suggests the potential for control of the Mn$^{2+}$ ion relaxation rate by means of the electric field in future experiments.

\end{abstract} 
\keywords{ODMR, QW, spin-lattice relaxation, carrier density}

\date{\today}
\maketitle

\section{INTRODUCTION}

The properties of the MBE grown nanostructures are determined by the complex interplay between the characteristics of used materials and growth conditions. In particular, the quality of the interfaces between the barrier and QW layer strongly impacts the optical and electrical quality of nanostructures.
The realistic transition between the barrier and the quantum well material differs from the idealized image of the perfectly-cut interface both regarding lateral and the growth direction. The main concern related to the latter issue is the intermixing, i.e., smoothing of the band profile due to unintentional migration of the chemical elements across the interface. Even in the case of negligible intermixing, the interface still can exhibit imperfection in the lateral direction in the form of a variation of the position of the interface. The variation of the position may be related to the single monolayer steps \cite{deveaud1984observation} but it can also have a larger extent for rough interfaces  \cite{Gaj_1994_PRB,regreny1987growth,grieshaber1994rough,Kossacki_1995_SSC}.

In the last case, irregularities of the interface introduce severe local deformation. Therefore magnetic ions near the interface experience different strain values than ions placed in the center of the QW. The distribution of the deformation of the crystal lattice at the sites of magnetic ions placed in various positions in nanostructure can affect the distribution of the ions' spin-lattice relaxation (SLR) time.

The magnetic ion relaxation can also be affected by interactions with carriers. The presence of the carrier gas increases the SLR rate \cite{Scherbakov_2001_PRB,Konig_2000_PRB}, due to effective energy transfer mediation between the ion and the lattice. In addition, the distribution of carriers may vary across the nanostructure. Consequently, a non-trivial SLR time distribution across the nanostructure may also be observed. 

All of the above shows that determining the detailed distribution of properties such as local deformation or distribution of carriers is essential for better design and
fabrication of devices combining electronic and spin effects. Moreover, this implies that one can utilize the control of the spatial distribution of the carrier gas density to adjust the properties of spin-lattice dynamics for spintronics applications.    

Studying such properties essentially requires a local approach, while techniques operating on mean values (such as Electron Paramagnetic Resonance -- EPR) are of limited use. Even the optical techniques, which are usually considered local, do not exhibit spatial resolution high enough to distinguish the signal originating from different depths of the QW.

\begin{figure*}
	\centering
		\includegraphics{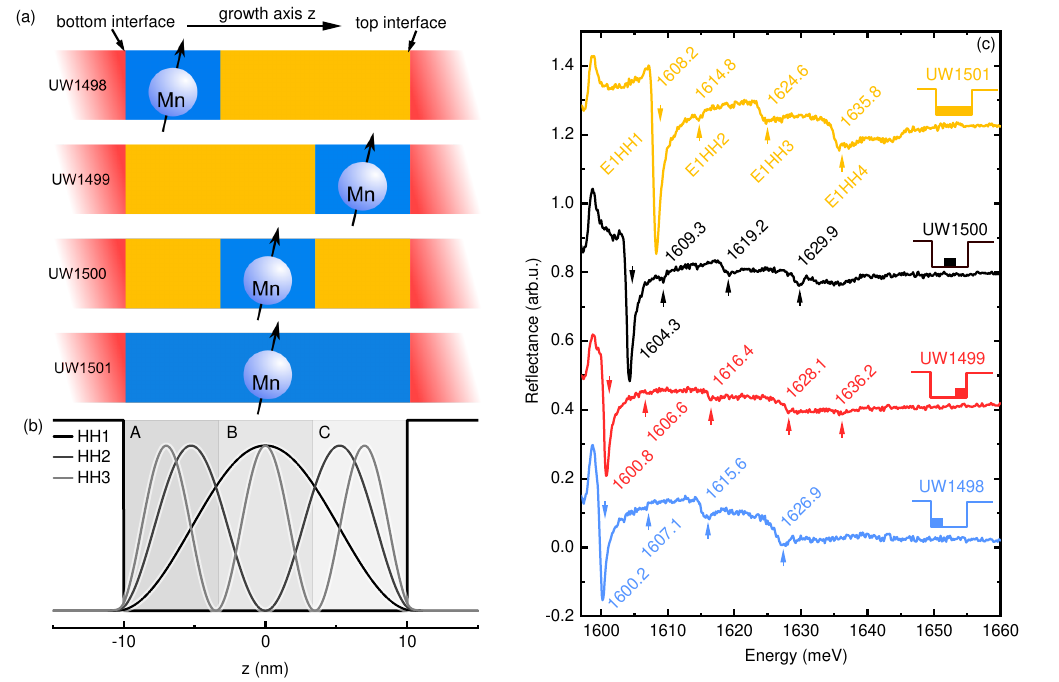}
		\caption{ (a) Schematic picture of the QWs samples used in this work. Manganese ions in the QW are distributed in various sections along the growth axis -- in either of the two sides of the QW near the interfaces with the barrier layer (samples UW1498 and UW1499), in the center of the well (UW1500), and in the entire QW (UW1501); (b) The schematic presentation of the spatial probability density of the first three heavy hole states in the model sample -- UW1501; (c) Reflectance spectra of studied samples measured at zero magnetic field.}
	\label{odbicia}
\end{figure*}

Here we exploit the fact that by tailoring the profile of the doping by the magnetic ions, we can restrict their presence to selected parts of the structure. The hyperfine structure of energy levels of each ion is sensitive to the deformation of the lattice at its site. 

Therefore, the deformation present in this part of the QW might be studied using the absorption of resonant microwave radiation with an applied magnetic field \cite{Lambe_1960_PR,wolos2004optical,Bogucki_2022_PRB}. This absorption can be detected either directly or indirectly by exploiting the fact that the optical properties of the studied material change when the paramagnetic resonance occurs. Thus, the resonance can be detected as a change of optical response, which is the essence of the optically detected magnetic resonance~(ODMR)~\cite{Gisbergen_1993_PRB,Ivanov_2001_APPA,ivanov2007time,Tolmachev_2020_N,Shornikova_2020_AN}. The ODMR technique allows studying observed effects locally in the area probed optically with micrometer resolution and only in the volume where the magnetic ions are introduced. The ODMR is particularly suitable for low-dimensional systems. Spatial selectivity, optical spatial resolution, and high sensitivity make ODMR a perfect technique for performing measurements on carriers and excitons coupled to the magnetic ions in nanostructures. The ODMR technique is especially useful for systems with large exchange interaction between magnetic ions and photocarriers -- a shining example of such a system is (Cd,Mn)Te in the diluted regime. More generally, the diluted magnetic semiconductors (DMS) exhibit the giant Zeeman effect \cite{Gaj_1994_PRB}, which connects the shift of the excitonic energy with the magnetization of the system of magnetic ions. 

The transitions between  Mn$^{2+}$ energy levels due to absorption of the microwave (MW) radiation leads to a decrease in magnetization (evidenced by the decrease of the giant Zeeman shift), which can also be described in terms of increased effective spin temperature.

In this work, we combine magnetooptical measurements and ODMR technique to study interfaces between (Cd,Mg)Te barriers and (Cd,Mn)Te QWs. In contrast to previous studies \cite{Gisbergen_1993_PRB,Ivanov_2001_APPA,ivanov2007time,Tolmachev_2020_N,Shornikova_2020_AN}, the optical detection in our ODMR experiment is based on reflectivity measurements, which allows us to monitor the behavior of not only the ground state exciton but also higher excited states.

The ODMR technique lets us examine the properties of the structures locally, in the volume where the magnetic ions are incorporated. This local approach, in conjunction with the unique design of the samples, is beneficial for studying the distribution of the deformation and SLR time along the growth axis of the QW. Restricting the Mn$^{2+}$ ion incorporation to the specific parts of the QW allows probing the mentioned properties locally -- near the interfaces and in the center of the well. By exploiting different excitonic states, we investigate other parts of the structure by varying the overlap with the magnetic ions. We also use time-resolved ODMR to study spin-lattice relaxation of the manganese ions incorporated in different positions along the growth axis.

\section{SAMPLES AND EXPERIMENTAL SETUP}

The samples containing single quantum wells were grown by molecular beam epitaxy on the semi-isolating GaAs substrate with 3.5\,$\mu$m CdTe buffer layer. The 20\,nm (Cd,Mn)Te QWs are surrounded by the (Cd,Mg)Te barriers with magnesium content of about 45\%. The barrier underneath the QW is 2\,$\mu$m thick (well above critical thickness), while the top barrier is 50\,nm. Manganese content is equal to about 0.5\%. This amount of Mn$^{2+}$ was chosen to assure sufficient giant Zeeman effect, but negligible direct ion-ion interactions and a small bandgap offset (less than 8\,meV) \cite{Gaj_2010_book}. The samples were designed so that manganese ions are placed at different positions along the growth axis as shown in~Fig.~\ref{odbicia}(a,b). In order to verify that all the produced samples actually follow the above layout, we characterized them using reflectance in a magnetic field. The samples are not intentionally doped. Nevertheless, the top barrier is only 50\,nm thick and due to the surface acceptors on (Cd,Mg)Te the nonintentional hole density in the QW is expected \cite{Maslana_2003_APL}. 

For the measurements, the samples were mounted on a holder designed specifically for ODMR experiments \cite{Patent_EN_2021}. The microwave radiation was provided to the samples with a microstrip antenna. The usage of an antenna has two main advantages over using a microwave cavity setup. First, the holder gives easy optical access both for PL and reflectance measurements, preserving the favored directions between the microwave magnetic and electric fields, as well as an external magnetic field. Second, it allows us to provide microwave radiation in a wide range of frequencies from 10\,GHz up to about 50\,GHz without any specific tuning or changing the cavity. All of the above allows measuring the paramagnetic resonance in a wide range of magnetic fields.

During measurements, the samples were immersed in pumped liquid helium (T$\approx$1.7\,K) in a magneto-optical cryostat with superconducting coils providing the magnetic field values up to 3\,T. The whole microwave setup was tested in a wide range of MW frequencies and amplitudes. The best antenna performance was selected by measuring the ODMR signal on a reference sample \cite{Bogucki_2022_PRB}. 

We used MW radiation and light illumination in the pulsed mode to avoid unwanted temperature drifts and other slow setup-related disturbances. The reflectance spectra were obtained using a filtered supercontinuum laser as a light source. The impinging light was chopped into pulses with an acousto-optic modulator (AOM) triggered by the signal of the MW generator (Agilent E8257D). The relative delay between MW and light pulses was varied in order to get temporal profiles of the ODMR signal. The width of pulses was a few ms for MW pulses and tens of $\mu$s for light pulses. The latter determines the resolution of transients. The pulse timing was controlled with a resolution of about 10\,ns. 
By keeping constant delays between the pulses, we performed time-integrated ODMR measurements.
The pulsed approach allowed us to correct for non-resonant thermal drift occurring during long experimental runs. At each data point, two cases were measured: with the light and MW excitation pulses in-phase (i.e. overlapping in time) and out-of-phase (i.e. with no overlap in time). The difference in the intensity of the signal (as defined in Sec. \ref{sec:odmr}) between these two situations gave us the ODMR signal robust against small temperature drifts.

\section{EXPERIMENTAL RESULTS}

\subsection{Magnetooptical measurements}

\begin{figure*}
    \centering
		\includegraphics{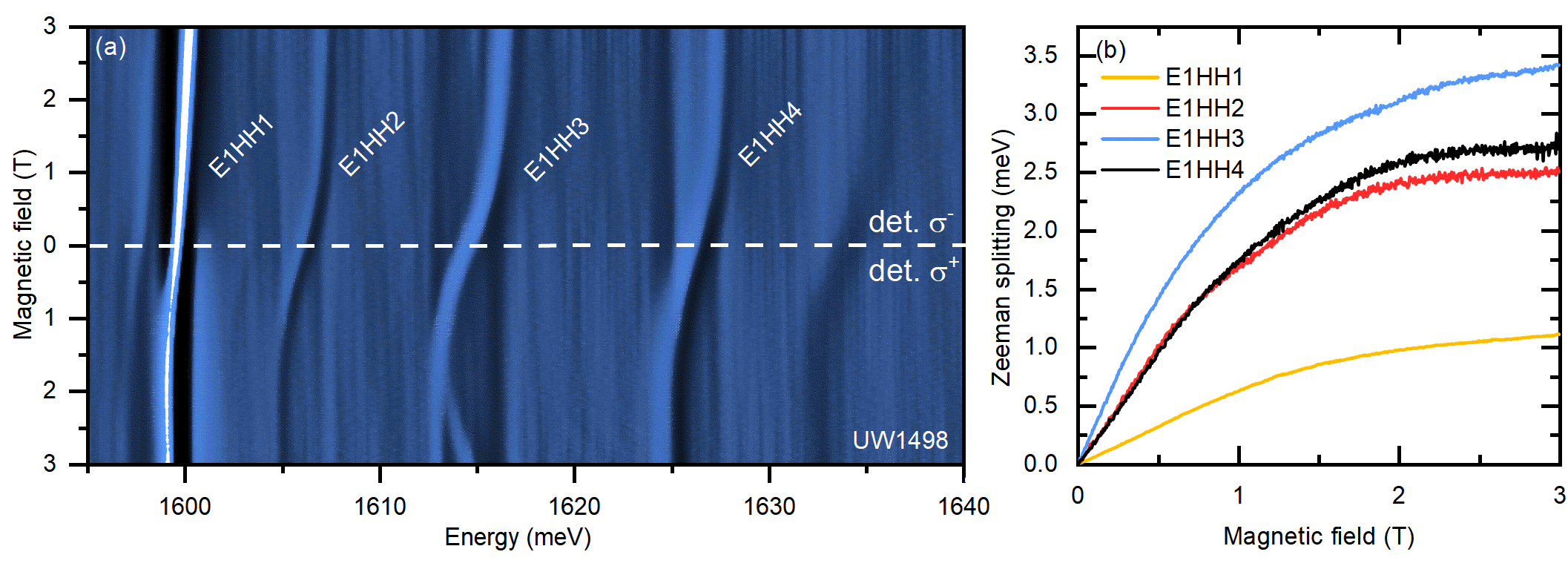}
		\caption{(a)Reflectance spectrum measured for different magnetic fields for sample UW1498; E1HHi denotes excitonic states, $\sigma^+/\sigma^-$ denotes detection polarization; (b) Giant Zeeman splitting of the different excitonic complexes observed for sample UW1498 vs. magnetic field}
	\label{UW1498_bez}
\end{figure*}

The magnesium content of the barrier layers was verified using standard reflectance measurements. We observed barrier exciton at around 2.6 eV, which corresponds to magnesium content equal to 46\%, as $E_\mathrm{g}^{\mathrm{Cd}_{1-x}\mathrm{Mg}_x\mathrm{Te}}(x) = Bx+E_\mathrm{g}^{\mathrm{CdTe}} = (1.85x+1.606) [\mathrm{eV}]$ \cite{Waag_1993_JoCG}. The value of manganese content was determined by fitting the modified Brillouin function \cite{Gaj_2010_book,Gaj_1994_PRB,gaj1979relation} with the parameters from ref. \onlinecite{Gaj_1994_PRB} to the exciton giant Zeeman shift obtained for the reference sample UW1501 (the Mn$^{2+}$ ions present in the whole QW). The Mn$^{2+}$ content was confirmed to be equal to 0.5\%.

The reflectance spectra for each studied QW exhibit several distinct features related to excitonic complexes of electrons and holes at different QW sub-bands, see~Fig.\ref{odbicia}(c). The most pronounced feature at the lowest energy, formed by the lowest energy electron and heavy-hole subbands, corresponds to the excitonic state denoted as E1HH1. The energy of this exciton is determined by several factors, such as the width of the QW (the same for all the samples), but also the overlap between the excitonic wavefunction and incorporated Mn$^{2+}$ ions, since the bandgap of (Cd,Mn)Te is larger than that of pure CdTe \cite{Gaj_2010_book}. For samples with magnetic ions placed only on the sides of the QW, the overlap between exciton wavefunction and magnetic ions gives small addition to ground state energy. In contrast, in the case of the sample with ions placed in the center of the QW, the overlap is significantly larger and results in a respectively larger increase of the ground state energy. Finally, manganese ions present in the whole well raise the QW transition energy by about 8\,meV, which is in agreement with 0.5\% Mn content determined from the giant Zeeman effect \cite{Gaj_2010_book}. Thus, excitonic states in all the samples are shifted towards higher energies in comparison to the QW without any Mn$^{2+}$ ions. For all samples, features corresponding to a number of excited states (E1HH$i$, $i=2-4$) are less pronounced but still visible. It is worth noting that we can observe the transition E1HH2, which should be forbidden in the perfectly symmetric QW. Its visibility suggests asymmetry of the wells' potential \cite{cibert1993piezoelectric,vanelle1996ultrafast}, observed even for the samples UW1500 and UW1501, which were designed with a symmetric Mn-doping profile. 

Figure~\ref{UW1498_bez}(a) shows an example map of the reflectance spectrum measured in the magnetic field applied in the Faraday configuration. Features corresponding to excitonic complexes -- ground state neutral exciton and a number of excited states -- are well-pronounced. The observed variety of the excitonic transitions exhibits different behavior in the magnetic field. The samples with the magnetic ions present on sides near the interfaces have a higher wavefunction-magnetic-ion overlap for higher states. This leads to a stronger coupling to the source of magnetization than for the ground state, causing a more pronounced Zeeman splitting, see~Fig.~\ref{UW1498_bez}(b). This result shows qualitative agreement with the model and further confirms that the manufactured samples are structured in accordance with our design. However, for the states higher than HH2, behavior in the magnetic field becomes more complicated as heavy holes are mixed with the light holes. The first light hole state is expected to be about 20\,meV above the HH1 state, due to the axial strain induced splitting \cite{Bogucki_2022_PRB} and different confinement energy. This splitting is caused by the deformation $\varepsilon_{\parallel}$ of the QW created by lattice mismatch with barrier \cite{Bogucki_2022_PRB}: $\Delta_{LHHH} = 2b (1+\frac{2C_{12}}{C_{11}}) \varepsilon_{\parallel} + \mathrm{const}$, where $b=-0.94$\,eV is shear deformation potential, $C_{11}, C_{12}$ are elastic stiffness constants and $\frac{2C_{12}}{C_{11}}|_{1.6K} = 1.39886$

\begin{figure}
    \centering
		\includegraphics{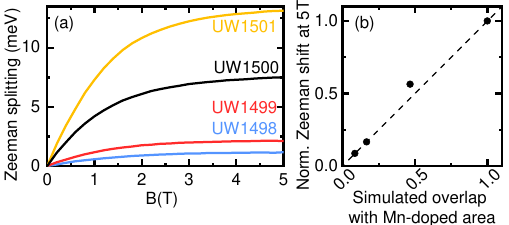}
		\caption{(a) Splitting in the magnetic field of the E1HH1 state measured for different samples at 1.7\,K (b) An overlap between the excitonic ground state and the magnetic ions volume obtained from the measurement of Zeeman shift vs. values calculated with consideration of electrostatic potential caused by the presence of the hole gas of density equal to $0.13\times10^{11}\,cm^{-2}$ with the spatial distribution given by ground heavy hole state probability.}
	\label{Zeeman_inter}
\end{figure}

Additionally, figure \ref{odbicia}(b) shows that samples UW1498 and UW1499 with Mn$^{2+}$ incorporated at two opposite interfaces have slightly different energies of E1HH1 exciton. They should be symmetric and have the same optical properties in the ideal case. However, the energy position of excitonic states differs here, which suggests a different overlap between carriers and Mn$^{2+}$ volume. Those differences are relatively small but clearly visible and become even more pronounced in the magnetic field. The Zeeman splitting of the exciton ground state in the sample UW1499 (with manganese ions near the top interface) is almost twice as high as splitting in sample UW1498 (bottom interface) see~Fig.\ref{Zeeman_inter}(a). Zeeman splittings for the UW1500 and UW1501 are even higher due to the higher wavefunction overlaps. 

Different overlap between magnetic ions volume and excitonic states derived both from different Zeeman splittings and zero field energies, along with the presence of E1HH2 transition, suggest the asymmetry of the QWs potential. At least two effects can explain the discrepancy between the Zeeman splittings in nominally symmetric samples: asymmetry of the top and bottom interfaces or built-in electric field. 

The first one corresponds to the non-abrupt, intermixed character of the interfaces. The formation process of interfaces depends on material properties and growth conditions. The QW asymmetry is caused by one of the intermixing mechanisms -- segregation: the growing material is dragged towards the outer layers, along the growth axis \cite{grieshaber1996magneto,Gaj_1994_PRB}. Thus, 
magnesium from barrier material is mixed into the QW material on the first interface and pushed out on the second one. This shift of composition moves potential in the first part of the well upwards, shifting the carrier wavefunctions away from the first interface. 

The second effect corresponds to unintentional p-type doping. Presented QWs are located near the surface of the sample (the top barrier is 50\,nm thick). In such a case, the surface states act as acceptors, causing the presence of hole gas in QW  \cite{Maslana_2003_APL}. However, in the reflectance spectrum, we do not observe any suggestion of charged excitonic features below the neutral exciton ground state energy. From that fact we can conclude that the carrier density in the QW, for studied samples is low -- i.e. estimated below 10$^{11}$\,cm$^{-2}$ \cite{kossacki1999neutral,kossacki2003optical,Kossacki_2004_PRB} -- yet still significant. The positive sign of charge carriers was confirmed using the spin-singlet triplet transition in low magnetic fields as in \cite{Lopion_2020_JEM,Kossacki_2004_PRB} (not shown). 

The presence of carrier gas in the QW (holes) and negatively charged states at the sample's surface result in an electrostatic field in the QW and the top barrier. This field modifies wavefunctions of carriers confined in QW. The modification is most pronounced for heavy holes due to their high effective mass. The hole gas builds a potential that drags holes wavefunctions towards the top barrier and electrons towards the bottom barrier. Consequently, wavefunction overlap with the manganese volume is higher in the sample with ions located near the top barrier, where the modified potential is deeper (UW1499).

We analyzed the plausibility of the above scenarios using numerical simulations. Therefore, nominal QW potential was calculated from composition using the same parameters as in Ref. \onlinecite{Bogucki_2022_PRB}. Moreover, modification of the potential was added based on the intermixed interfaces or on the electric field, which is rooted in carrier distribution along the z-axis of the sample (holes in the QW and electrons on the surface). The calculation of the wavefunctions perturbated by the electric field were performed with the use of a single-step Schrodinger-Poisson Solver, which is a sufficient approach for low carrier densities. We calculated overlaps between wavefunctions and the magnetic ion volumes for each sample as integrals over the proper part of the QW (A,B,C or the whole QW) of the calculated wavefunctions density.

We found that the change of the potential corresponding to the intermixed interfaces alone is insufficient to explain observed differences in the Zeeman splitting when taking reasonable intermixing coefficients below 1\,nm \cite{Gaj_1994_PRB}. At the same time, calculations for the scenario with the electric field involved show excellent agreement with the measured data when assuming carrier density equal to 0.13$\times$10$^{11}$\,cm$^{-2}$. In the first approximation we assumed that carrier density is the same for all the samples. That assumption seems to be reasonable as the design of the samples is very similar -- especially they have the same thickness and composition of the top barrier layer but generally the carrier density can slightly vary between the samples. Nontheless from the very good agreement for the samples UW1498 and UW1499 we conclude that for both of the samples the carrier density is very close to the one used in the calculation. The noteworthy is the fact that the Zeeman splitting obtained for the sample UW1500 (the one with the manganese ions placed in the middle) is slightly higher than it should be in the ideal, symmetric case. That cannot be obtained with the change of the carrier density as the addition of built-in electric field can only decrease the overlap between the ground state wavefunctions and the Mn ions volume in the centre of the QW.

The calculated overlap was defined as integral of state probability over the part of the QW where Mn$^{2+}$ ions are present. Assuming 100\% overlap between HH1 wavefunction and magnetic layer for sample UW1501, which causes a Zeeman splitting equal to 13.2\,meV in 3\,T, we can calculate the overlap between HH1 and magnetic layer for the other samples. This overlap is obtained as a ratio of experimentally determined Zeeman splitting for the analyzed sample and the sample UW1501, see~Fig.~\ref{Zeeman_inter}(b). We can compare those values with the HH1 -- Mn$^{2+}$ overlap. As for sample UW1501, where manganese ions are present in the entire well, this overlap is equal to almost 1. For other samples, the overlap is respectively lower. Finally, we conclude that observed differences of giant Zeeman shift between samples UW1499 and UW1498 correspond to the electric potential built in the samples. 

\subsection{ODMR measurements \label{sec:odmr}}

\begin{figure*}
    \centering
		\includegraphics{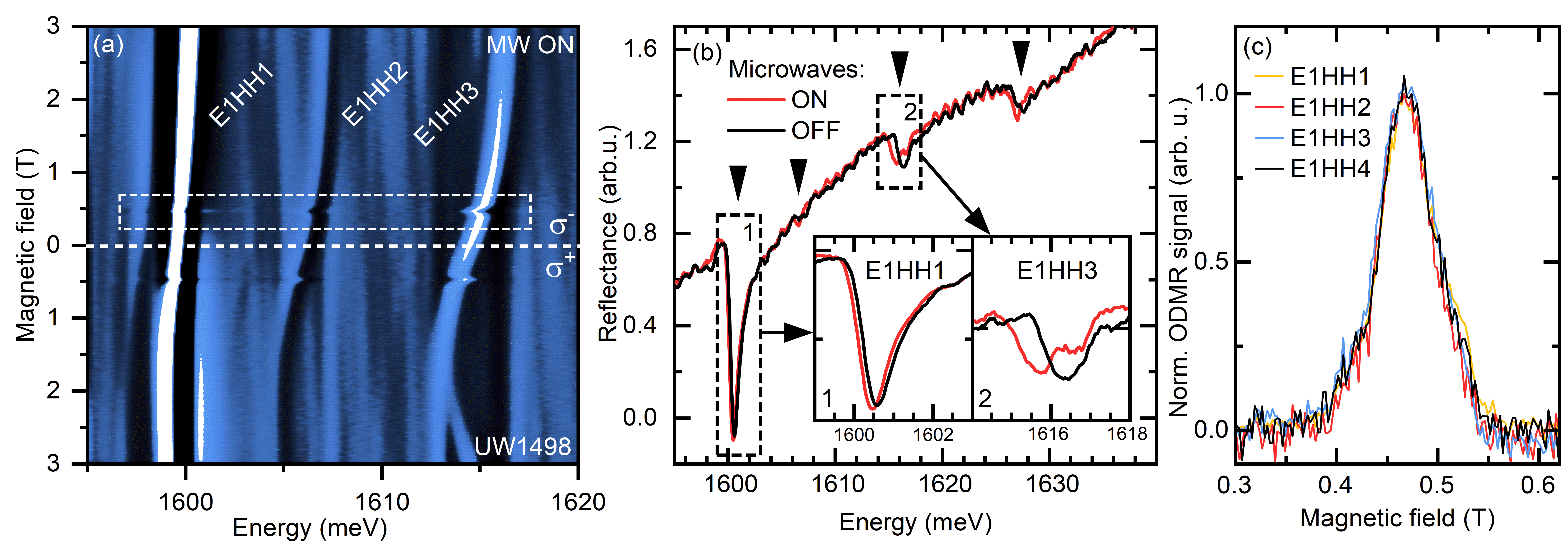}
		\caption{(a) Reflectance spectrum measured for the different magnetic fields in the presence of microwave radiation. In the resonant magnetic field (here 0.57\,T for 15.85\,GHz) the energy shift is observed for all excitonic complexes -- (e.g., see dashed rectangle); (b) Reflectance spectrum collected in magnetic field  0.57\,T with resonant microwave radiation (15.85\,GHz) with pulses in positions ON and OFF; insets represent two different excitonic states -- ground state E1HH1 and one of the excited states E1HH3; (c) Normalized ODMR signal measured for all visible excitonic complexes of UW1498 sample.}
	\label{UW1498_ref_zMW}
\end{figure*}

Figure~\ref{UW1498_ref_zMW}(a) shows the reflectance spectra measured versus the magnetic field under microwave radiation. All the excitonic lines in the reflectance spectrum exhibit sensitivity to the applied microwave radiation for magnetic resonance fields (near $\pm$0.5~T). We observe that in the magnetic resonance fields, all the features are shifted towards the energies corresponding to lower absolute values of the magnetic field, see a dashed rectangle in Fig.~\ref{UW1498_ref_zMW}(a). Furthermore, the shift of the excited states under microwave radiation is more pronounced than the shift of the ground state, see~Fig.~\ref{UW1498_ref_zMW}(b). That effect corresponds to a larger change of the Zeeman shift for excited complexes.

\begin{figure}
	\centering
		\includegraphics{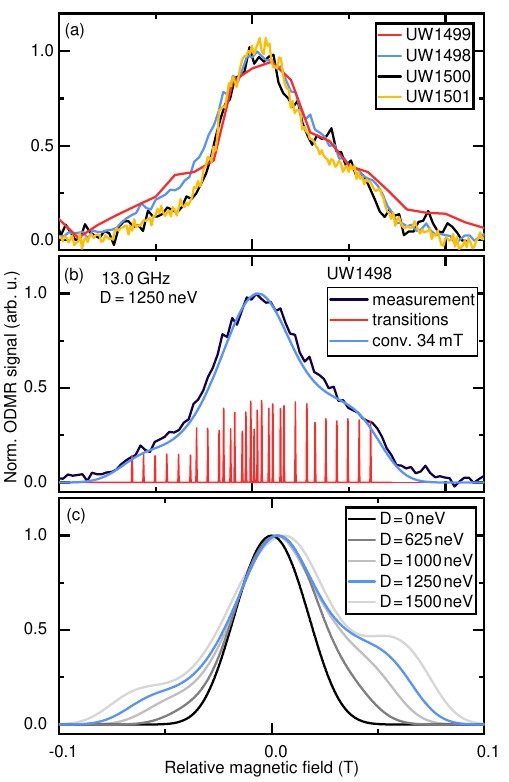}
		\caption{(a) Normalized ODMR signal vs. magnetic field extracted from E1HH1 state for all measured samples (b) ODMR signal vs. magnetic field measured for the UW1498 QW with the simulated ODMR signal. The red spiked curve shows calculated magnetic transitions without any line broadening. The blue curve is a result of the convolution of the red-lined spectrum with the Gaussian kernel. (c) Compilation of the simulated shapes of ODMR signal as a function of magnetic field for different values of the spin Hamiltonian D parameter.}
	\label{shape}
\end{figure}

We define the ODMR signal as the difference in the energy position of the excitonic line in the reflectance spectrum for microwave and light pulses overlapping (ON) and shifted (OFF). An example of a representative ODMR signal as a function of magnetic field for fixed microwave frequency is presented in~fig.~\ref{UW1498_ref_zMW}(c). The shape of the signal versus magnetic field obtained for all the excitonic states visible in the single QW is the same, see Fig.~\ref{shape}(a). This fact clearly shows that all the complexes are probing the ensemble of the magnetic ions with the same properties. 

We also observe that the shape of the ODMR signal obtained for different samples is similar, see Fig.~\ref{shape}(a).
The main factor that determines the overall shape of the ODMR signal as a function of the magnetic field at low temperatures is the deformation of the crystal lattice surrounding probed magnetic ions,~\cite{Bogucki_2022_PRB}. The lattice mismatch between the layers of the structures results in nonzero strain-induced axial-symmetry spin parameter D of the Hamiltonian of the Mn$^{2+}$ ion \cite{Qazzaz_1995_SSC}. Angular resolved OMDR measurements can determine the value of the spin Hamiltonian parameter D. In the previous work \onlinecite{Bogucki_2022_PRB} we assumed homogenous strain within the QW, and we probed whole of its volume with preferential sensitivity of the central region. In this work we test this assumption by studying QW with Mn$^{2+}$ restricted to the specific parts along axis growth, thus with preferential sensitivity in different parts of the QW. 

As the bottom barrier layer is 2\,$\mu$m thick we assume that it is relaxed on the interface with the QW and it governs the lattice mismatch for the QW layer. The nominal deformation coming from this lattice mismatch is $\varepsilon_\parallel = \frac{a_{barrier}}{a_{QW}}-1$ where $a_{barrier}$ and $a_{QW}$ are lattice constants for barrier and QW layers, respectively. In the case of (Cd,Mg)Te/(Cd,Mn)Te QW with Mg content about 46\% and Mn content about 0.5\% the lattice constant are $a_{barrier}=6.45248$\,\AA~and~$a_{QW}=6.48028$\,\AA~deformation is  $\varepsilon^{300K}_\parallel = -4.195\,\permil $. Taking into account the different thermal expansion coefficients of CdTe and substrate GaAs the deformation for measurement temperature is $\varepsilon^{1.6K}_\parallel = -4.837\,\permil$. As it was presented in \onlinecite{Bogucki_2022_PRB}, following the equation $D = -\frac{3}{2} G_{11} \left (1+\frac{2C_{12}}{C_{11}}\right) \varepsilon_\parallel$ such value corresponds to the D parameter about 1250\,neV. All the constants are the same values as presented by Bogucki et al. in Ref.~\onlinecite{Bogucki_2022_PRB}. 

We compare measured ODMR spectra with the numerically obtained positions and intensities of absorption lines based on the Mn$^{2+}$ spin Hamiltonian like in ref. \cite{Bogucki_2022_PRB}. The finite linewidth is obtained using the phenomenological broadening approach. In figure~\ref{shape}(b) experimental ODMR signal vs. magnetic field measured for the sample UW1498 is shown along with the calculated one. Each of the 30 transition lines (marked in red) is convoluted with a gaussian kernel of FWHM= 34\,mT (blue). We satisfactorily reproduce the measured ODMR spectra for a D parameter value of 1250\,neV which is in a good agreement with the strain nominally present in the structure.

Mention worthy is the sensitivity of the shape of the ODMR signal vs. magnetic field to changes of the D parameter as is presented in~figure~\ref{shape}(c). The agreement with the D corresponding to the nominal lattice mismatch suggests that the strain is distributed equally along the growth axis of the QW layer for each of the samples.

\begin{figure}
	\centering
		\includegraphics{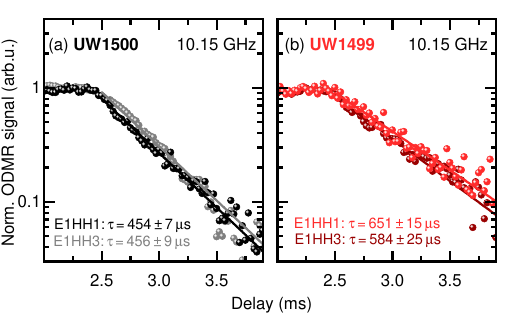}
		\caption{Normalized time-resolved ODMR signal for E1HH1 and E1HH3 excitonic states for samples: (a) UW1500 and (b) UW1499 for resonant frequency f=10.15~GHz. The observed ODMR amplitude relaxation time is the same regardless of the excitonic state used for the probe.}
	\label{delaye1}
\end{figure}

\begin{figure}
	\centering
		\includegraphics{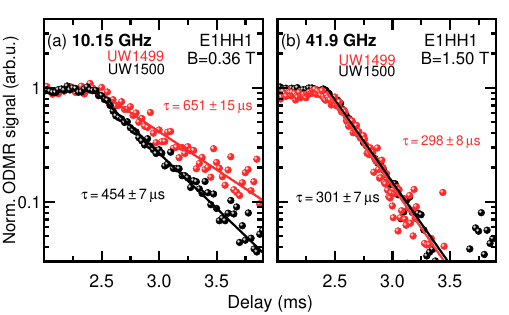}
		\caption{Time-resolved ODMR signal amplitude for samples UW1499 and UW1500 extracted from E1HH1 state for different resonance frequencies: (a) 10.15\,GHz and (b) 41.9\,GHz.}
	\label{delaye2}
\end{figure}

\subsection{Time-resolved ODMR measurements}

We have performed time-resolved ODMR measurements on each of the presented samples and the visible excitonic complexes. The spin-lattice relaxation (SLR) time measured on the higher excitonic states has the same value as those obtained on the ground state. The example of the temporal profiles is shown in~Fig.~\ref{delaye1}(a,b). The SLR time measured on the excited states is the same as for the ground state for all the samples. 

As expected, the SLR time depends on the magnetic field (resonant MW frequency), see~Fig.~\ref{delaye2}(a,b). We observe a significant shortening of the SLR time for higher magnetic fields, where the differences between samples become imperceptible.  

The analysis of measured SLR times as a function of magnetic field for different samples reveals non-trivial effects. In~figure~\ref{delaye2}(a), temporal profiles measured for 10.15\,GHz resonance for samples UW1500 and UW1499 are compared. This frequency corresponds to the paramagnetic resonance in a magnetic field equal to approximately 0.36\,T. In this case, we observe a significant difference in the SLR times. The situation is noticeably different when it comes to measurements in higher magnetic fields -- temporal profiles obtained for 41.9\,GHz resonance (about 1.5\,T) are shown in~Fig.~\ref{delaye2}(b). Here, the SLR times are the same for both samples.   

\begin{figure}
	\centering
		\includegraphics{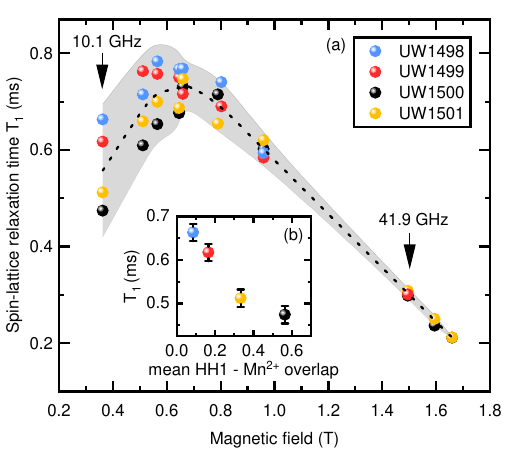}
		\caption{(a)SLR time versus the magnetic field. Black arrows mark magnetic fields for which temporal profiles are presented in~Fig.\ref{delaye2}(a) and (b); shadowed area in the background is a guide for the eye. As the magnetic field increases, variation of the SLR times between the samples decreases; (b) mean HH1 -- Mn$^{2+}$ overlap per manganese volume vs. SLR time for $B=0.36$\,T}
	\label{gorycogram}
\end{figure}

A summary of the measured SLR times for all samples and the magnetic fields is presented in~figure~\ref{gorycogram}(a). For all analyzed samples, a maximum value of SLR time is observed around 0.6\,T. The SLR times decrease with the magnetic field and follow the same curve for all the samples for higher magnetic fields. However, the measurements at lower magnetic fields show notable differences for different QWs. Especially the samples with Mn$^{2+}$ ions placed in the center of the QW (UW1500 and UW1501) have significantly shorter SLR times than the samples with ions placed on the QW interfaces. This finding suggests an essential role of the Mn$^{2+}$ -- HH1 probability density overlap in the relaxation mechanisms at lower magnetic fields. 

\section{Discussion of spin relaxation}

What we observe is that SLR depends on the applied magnetic field following the curve shown in figure 8(a). The observed change of behavior of the SLR time in the magnetic field suggests at least two regimes with different leading mechanisms causing SLR relaxation.

In general, spin dynamics in DMS are controlled by the energy and spin transfer between the Mn$^{2+}$ system, carriers, and the crystal lattice. In our experiment, we heat up the Mn$^{2+}$ system resonantly with the MW radiation. The Mn$^{2+}$ system can relax directly – with spin and energy dissipated to the phonon bath or indirectly via carriers present in the structure. Measured SLR is a result of those two paths. The rates of these energy transfers depends on the effective temperatures of the systems and densities of states available for scattering.

\begin{equation}
    \frac{1}{\tau_\mathrm{SL}} = \frac{1}{\tau^0_\mathrm{SL}} + \frac{1}{\tau_{h-Mn}}
\end{equation}

We denote here the base SLR time of Mn ions spins and energy directly to the lattice as $\tau^0_\mathrm{SL}$. In the regime of the high magnetic fields, this mechanism takes the leading role as its rate is determined by the increasing density of phonon states allowed for scattering at the energy corresponding to the Zeeman splitting. It is well established that the magnetic field dependence of this mechanism is power of B: $1/\tau^0_\mathrm{SL} \sim B^n $ \cite{Strutz_1992_PRL,strutz1993principles,Goryca_2015_PRB}. The exponent of B depends on the model applied for calculating the density of phonon states and the type of scattering mechanism (direct or indirect). Such increase of the SLR rate of Mn$^{2+}$ with the magnetic field was studied for manganese ions in different systems: bulk crystals, quantum wells, and quantum dots \cite{Scherbakov_2001_PRB,Strutz_1992_PRL,Goryca_2015_PRB,strutz1993principles}. In our particular case, for $B>1$\,T we find that the relaxation time is shorter upon increasing the magnetic field what is in the agreement with previous reports. In this regime all studied samples exhibit the same SLR time, which is expected due to the same phonon spectrum.

Upon approaching the low field limit, the phonon based mechanism is less and less efficient, and is surpassed by another mechanism: spin and energy scattering via the carrier's system. The ultimate intrinsic limitation for the capacity of this channel is given by the carrier spin lifetime $\tau_\mathrm{S}$, which is very short \cite{Scherbakov_2001_PRB,akimov2006multiple}. As the $\tau_\mathrm{S}$ is very short, in practical situations this channel is limited by the carrier gas-Mn system scattering time $\tau_\mathrm{h-Mn}$. 

Conceptually, the simplest mechanism of energy transfer between Mn and carrier system is the spin flip-flop \cite{Konig_2000_PRB,Scherbakov_2001_PRB,akimov2006multiple}, in which a Mn$^{2+}$ ion and a carrier undergo a simultaneous spin flips. The energy conservation requires that the two involved states of the carrier are separated by the energy equal to the Zeeman splitting of the Mn$^{2+}$ ion. Since the initial and final carrier states correspond to different spin subbands, the availability of the populated initial states crucially depends on the Fermi level $\varepsilon_\mathrm{F}$ (and thus the carrier density). The giant Zeeman splitting of the valence band is equal $E_\mathrm{z,c}$ = $\mu_\mathrm{B} g^\mathrm{eff}_\mathrm{c} B$, with effective g-factor $g^\mathrm{eff}_\mathrm{c}$, which is extraordinarily big -- i.e. for sample UW1501 $g^\mathrm{eff}_\mathrm{h} \sim 110 $. Due to the large value of the $g^\mathrm{eff}_\mathrm{c}$, the splitting $E_\mathrm{z,c}$ exceeds $\varepsilon_\mathrm{F}$ even in relatively small magnetic fields. In such a case, only the lower spin subband is occupied, while the upper band is empty. As the Zeeman splitting of Mn$^{2+}$ ions ($E_\mathrm{Z,Mn}$)  is relatively small (g-factor $= 2$), the energy transfer from magnetic ions to the carriers becomes impossible, and the $\tau_\mathrm{h-Mn}$ rate decrease with the increase of the magnetic field. Different Zeeman splittings of Mn$^{2+}$ and carriers leads to decoupling of both systems at higher magnetic fields. The maximum of the SLR via carrier channel is in zero magnetic field, while the separation between energy levels of Mn$^{2+}$ ions and spin subbands are similar. For $B>0$, the SLR rate via this channel drops due to the detuning of the energy levels of those two systems. As the carriers sub-system has a finite effective temperature also the states with higher k-vectors are occupied. Those states can take part in the relaxation even if the separation of sub-bands in the magnetic field exceeds $E_{z,Mn}$ in the magnetic field. A very similar situation is when the carrier density is sufficiently large, and the Fermi level exceeds the $E_\mathrm{Z,c}$. The flip-flop interactions model leads to the $\tau_\mathrm{h-Mn}$ dependence on the magnetic field $B$, carrier density via Fermi level $\varepsilon_\mathrm{F}$, carriers' spins scattering time $\tau_\mathrm{S}$ and effective temperatures for both subsystems -- magnetic ions $\beta_\mathrm{Mn}$ and carriers $\beta_\mathrm{c}$ \cite{Scherbakov_2001_PRB,Konig_2000_PRB}: $\tau_\mathrm{h-Mn}= \tau (B, \varepsilon_\mathrm{F}, \tau_\mathrm{S}, \beta_\mathrm{c}, \beta_\mathrm{Mn})$.

Aforementioned mechanism is present both in p- and n-doped systems. In comparison, the holes have a higher effective mass $m_{\mathrm{eff}}$ than electrons, causing Fermi level for the density of holes $n$ to be lower than for the same density of electrons $n$ ($\varepsilon_\mathrm{F} = \frac{n \pi \hbar^2}{m_\mathrm{eff}}$). Nevertheless, the interactions with holes can provide more efficient relaxation channel \cite{akimov2006multiple,scherbakov2001spin}. Holes have a stronger effect on the SLR as they have a larger density of states (due to the larger effective mass), and (p-d) exchange interaction of Mn$^{2+}$ ions with holes is stronger than with electrons \cite{Gaj_2010_book}. Moreover, there may also be a second process that allows SLR of Mn$^{2+}$ ions via hole gas. It was considered by Akimov et al. \cite{akimov2006multiple} that while for the electrons there are only flip-flop transitions, for the holes, the transfer of spin may be possible during the relaxation in a single Zeeman subband due to the mixing of heavy and light hole states, proportional to $\sim k^2$. This mechanism might be an even more effective relaxation channel than flip-flop transitions if Fermi level is too low for the latter one.

Summing up: observed drop of the SLR time in low magnetic fields may be explained by the interactions with hole gas. Moreover, for the magnetic field below 0.6\,T, we observe the differences in SLR times between the studied samples. Those differences are correlated with the mean overlap between the manganese ions and the local wavefunction density of HH1, which corresponds to carrier gas distribution, see Fig. 8(b). The observed differences between studied samples could be explained in the carrier gas density and ion-carriers interactions frame. The magnetic ions incorporated in the center of the UW1500 sample have a higher overlap with carrier gas wavefunctions than the ions on the sides of UW1499 or UW1498. Possibly, that makes the sample UW1500 more sensitive to the effects corresponding to the free carriers than the other samples. Moreover, the asymmetry between samples UW1499 and UW1498 discussed in the previous part can also affect the SLR time in the way we observe it. Namely, the relaxation for the UW1498 is slower – as the overlap of the holes and ions for this sample is lower due to asymmetry caused by the electric field.

\section{Conclusions}
Our work shows that the ODMR technique is a powerful tool for studying the distribution of the strain in the nanostructures along the growth axis. It also reveals interactions between magnetic ions and charge carriers. Both strain and interactions with carriers can affect SLR, which may be essential for future applications. With the ODMR technique, we performed measurements on the series of samples with Mn$^{2+}$ ions incorporated in the different regions of the QWs -- on the sides -- near the barrier layers, and in the center. In addition, we found that we can probe the different layers of the well along the growth axis using the excited excitonic states.

We do not see any indications that the deformation of the crystal lattice is changing along the growth axis. Therefore, we conclude that the growth of the QW layers is homogeneous.  

Additionally, the carriers play an essential role in the presented samples. Even if the hole density in the QWs seems to be low, the introduced electric field affects the shape of the confined wavefunctions and results in a consecutive change of their overlap with magnetic ions. Effects related to free charges are more pronounced for low magnetic fields -- which is in agreement with the previous studies \cite{Scherbakov_2001_PRB, Konig_2000_PRB}. We conclude that the overlap between magnetic ions and hole gas is the main factor varying across this series of QWs.  Of special interest is the case of assymetrically doped QWs, which evidence the importance of the electric field across the structure. Our observations prove the potential to cotrol the SLR time e.g., by deterministic tuning of the electric field in gated samples. 

\section{Acknowledgements}
This work was supported by the Polish National
Science Centre under decisions DEC-2016/23/B/ST3/03437,~DEC-2015/18/E/ST3/00559,~DEC-2020/38/E/ST3/00364,~DEC-2020/39/B/ST3/03251.

\end{document}